\documentclass[pdflatex,default]{sn-jnl_ARXIV}
\jyear{2022}%

\usepackage{float}
\usepackage{amsmath}
\usepackage{comment}
\usepackage{epstopdf}

\usepackage{siunitx}
\usepackage{caption}
\usepackage{calc}
\usepackage{rotating}
\usepackage{xcolor}
\usepackage{amssymb}
\usepackage{listings}
\newcommand*\circled[1]{\tikz[baseline=(char.base)]{
            \node[shape=circle,draw,fill=white,inner sep=2pt] (char) {#1};}}

\usepackage{subcaption}
\usepackage{mwe}

\usepackage{hyperref} 

\usepackage{pifont} 

\usepackage{tablefootnote}

\usepackage{subcaption}

\normalbaroutside
\usepackage{tikz}
\usepackage{pgfplots}
\usepackage{glossaries}
\newacronym{bc}{BC}{boundary condition}
\newacronym[longplural={degrees of freedom}]{dof}{dof}{degree of freedom}
\newacronym{fem}{FEM}{finite element method}
\newacronym{fe}{FE}{finite element}
\newacronym{ga}{GA}{genetic algorithm}
\newacronym{nsga}{NSGA}{non-dominated sorting genetic algorithm}
\newacronym{moo}{MOO}{multi-objective optimization}

\newcommand{\commentDL}[1]{\textbf{\textcolor{orange}{#1}}}

\begin{document}

\title[Quadrupole Magnet Design based on Genetic Multi-Objective Optimization]{{Quadrupole Magnet Design based on Genetic Multi-Objective Optimization}}

\author{\fnm{Eric} \sur{Diehl}}
\author*{\fnm{*Moritz} \sur{von Tresckow}}\email{moritz.von\_tresckow@tu-darmstadt.de}
\author{\fnm{Lou} \sur{Scholtissek}}
\author{\fnm{Dimitrios} \sur{Loukrezis}}
\author{\fnm{Nicolas} \sur{Marsic}}
\author{\fnm{Wolfgang F. O.} \sur{Müller}}
\author{\fnm{Herbert} \sur{De Gersem}}

\affil[]{\orgdiv{Institute for Accelerator Science and Electromagnetic Fields (TEMF)}, \orgname{Technische Universit\"at Darmstadt}, \orgaddress{\street{Schlossgartenstr. 8}, \city{Darmstadt}, \postcode{64289}, \country{Germany}}}

\abstract{
This work suggests to optimize the geometry of a quadrupole magnet by means of a genetic algorithm adapted to solve multi-objective optimization problems. 
To that end, a non-domination sorting genetic algorithm known as NSGA-III is used.
The optimization objectives are chosen such that a high magnetic field quality in the aperture of the magnet is guaranteed, while simultaneously the magnet design remains cost-efficient. 
The field quality is computed using a magnetostatic finite element model of the quadrupole, the results of which are post-processed and integrated into the optimization algorithm. 
An extensive analysis of the optimization results is performed, including Pareto front movements and identification of best designs.
}

\keywords{Magnet Design, Quadrupole, Geometry Optimization, Multi-Objective Optimization, Genetic Algorithm, NSGA-III}

\maketitle

\section{Introduction}\label{sec:Introduction}
Multipole magnets are essential components of particle accelerators and are of paramount importance for the success of particle physics experiments that are performed in accelerator facilities \cite{russenschuck2011field}. 
Quadrupole magnets, in particular, are crucial for keeping particle beams focused on their desired trajectories \cite{Conte1991Intro}.
The very nature of particle physics experiments places very high demands on the operation and performance of accelerator magnets.
Therefore, their production must be based on meticulous design procedures.  

Focusing on the case of quadrupole magnet design, common practice dictates to formulate the design objectives into an optimization problem with respect to the geometric parameters of the quadrupole, while certain operational parameters must also be taken into account during the optimization process. 
Therein, the electromagnetic phenomena taking place inside the magnet are typically simulated using a digital magnet model and are included into the optimization as performance measures \cite{Ahmad2020Actuator, Kalimov2014Pole, Sempere2021optimisation, ion2018robust, pels2015optimization}.
However, such geometry optimization problems are notoriously hard to solve, in fact, NP-hard \cite{Arora03approximationschemes}, due to the fact that their computational complexity increases exponentially with the number of solution candidates.
In many cases, the only viable option is to employ search algorithms that are capable of comprehensively exploring the parameter space and finding an adequate configuration at a reasonable computational cost.
Popular methods of choice are Monte Carlo algorithms \cite{MAVROTAS2015193, Mikhailov1992Optimization, BARNOON20222747}, simulated annealing \cite{Donnelly1987Geometry, Mundim1996Geometry} or \glspl{ga} \cite{hou1994genetic,schmitt2001theory, weile1997genetic, KONAK2006992}, all of which have been successfully applied to a number of geometry optimization problems \cite{asselineau2015integration, uler1994utilizing,simkin1992optimizing}. 

This work focuses on the use of \glspl{ga} for the purpose of quadrupole design.
Originally, \glspl{ga} have been developed for unconstrained, single-objective optimization problems, which constitute their natural domain of application \cite{Mahrach2020Comparison}. 
However,  most geometry optimization applications, also including quadrupole magnet design, feature multiple and often conflicting objectives. 
In such cases, a \gls{moo} \cite{gunantara2018review, justesen2009multi} must be solved instead. 
Therein, the goal is to find so-called \emph{Pareto optimal} solutions, which cannot be further improved with respect to one objective without worsening another \cite{luc2008pareto}.
Such \Gls{moo} problems are faced and tackled increasingly more often in various engineering applications that concern geometry optimization \cite{yadav2020geometric, di2020many, garcia2022multiobjective, bizzozero2021multi}.
To address these problems, traditional \glspl{ga} have been extended, e.g., in the form of so-called \glspl{nsga} \cite{NSGAI, NSGAII, NSGAIII, jain2013evolutionary}.

Regarding the quadrupole magnet design problem considered in this work, the objectives of the \gls{moo} are: (a) the maximization of magnetic field quality within the aperture of the quadrupole magnet, and (b) the minimization of the magnet's radius.
The former objective is aligned with the desired magnet operation with respect to beam focusing.
The latter objective aims at designs that use the minimum necessary amount of magnet material, so that production costs remain acceptable. 
The magnetic field quality is first computed using a magnetostatic \gls{fe} magnet model and then expressed in the form of relative multipole coefficients on a reference radius. 
The evaluation of the relative multipole coefficients is integrated into the fitness function of an \gls{nsga}, such that only solutions which maximize the magnetic field quality are allowed to propagate through the evolution process, as dictated by the first optimization objective.
In particular, the algorithm known as \gls{nsga}-III \cite{NSGAIII} is employed.
The solutions are further constrained by the second objective, i.e., the need for a minimal magnet radius.
The Pareto-optimal solutions, i.e., those that provide an acceptable balance with respect to both objectives, are then analyzed such that the best magnet design candidates are identified.

While works featuring \gls{ga}-based \gls{moo} algorithms for the optimization of accelerator systems do appear in the literature \cite{edelen2020machine, hofler2013innovative, husain2018constrained, korchuganov2018multiobjective, neveu2019parallel, yang2009global}, the application of such optimization algorithms to accelerator magnet design seems to have been so far neglected. In fact, the only relevant work that the authors are aware of is \cite{russenschuck2011field}, which mentions the option of optimizing accelerator magnets by means of \gls{moo} based on \glspl{ga}, however, without providing any verification in the form of numerical experiments. The present work aims to fill this gap.

The rest of this paper is organized as follows. Section~\ref{sec:ProblemFormulation} presents the problem formulation with respect to computing the magnetic field of the magnet for a given design, the magnet model considered in this work, and the computation of the magnetic field quality in the aperture of the magnet. Section~\ref{sec:Optimization} presents the general formulation of \gls{moo} problems and discusses \glspl{ga} suitable for the solution of such problems, in particular the \gls{nsga}-III. In Section~\ref{sec:magnet-optimization}, the \gls{moo} problem with respect to the quadrupole magnet design is formulated. The numerical results obtained by means of the \gls{nsga}-III algorithm are presented and extensively discussed in Section~\ref{sec:Results}. Finally, conclusions are drawn in Section~\ref{sec:ConclusionAndOutlook}.

\section{Problem Formulation and Magnet Model} \label{sec:ProblemFormulation}

The physical behavior of a quadrupole magnet can be fully described by the Maxwell equations, which provide the mathematical foundation for classical electrodynamics.
However, in most cases, it suffices to provide an approximate description of the electromagnetic phenomena appearing in a given problem setting, thus simplifying the underlying equations and the resulting simulation model. 
For our particular application of a quadrupole magnet with nonlinear materials, it is sufficient to consider a magnetostatic representation of the underlying magnetic field quantities. Therefore, we employ the magnetic vector potential formulation in the steady state regime. 

Considering the 3D case, the magnetic vector potential formulation is given as
\begin{equation}
    \nabla \times (\nu(\mathbf{b}) \nabla \times \mathbf{a}) = \mathbf{j}, \label{eq:MS1_3D}
\end{equation}
where $\nu$ is the magnetic reluctivity tensor, $\mathbf{b}$ the magnetic flux density, $\mathbf{a}$ the magnetic vector potential, and $\mathbf{j}$ the current density. 
Equation \eqref{eq:MS1_3D} is discretized by employing a projection on \eqref{eq:MS1_3D} with a test function $\mathbf{v}$ on a computational domain $\Omega$ with boundary $\partial\Omega = \partial\Omega_\mathrm{N} \cup \partial\Omega_\mathrm{D}$, where $\partial\Omega_\mathrm{N}$ and $\partial\Omega_\mathrm{D}$ respectively denote the boundary parts where Neumann and Dirichlet \glspl{bc} are imposed \cite{bossavit1998computational}.
The corresponding weak formulation reads: 
\begin{align}
    &\mathrm{find} \ \mathbf{a} \in \mathbf{L}_0(\mathrm{curl},\Omega) \ \text{such that} \nonumber
    \\ 
    &\int_\Omega \nu(\mathbf{b}) \nabla\times\mathbf{a}\cdot\nabla\times\mathbf{v}\, \mathrm{d}\Omega = \int_\Omega \mathbf{j}\cdot\mathbf{v} \, \mathrm{d}\Omega, 
    \label{eq:weakFormulation}
\end{align}
for all test functions $\mathbf{v}$ in the space $\mathbf{L}_0(\mathrm{curl},\Omega)$, defined as 
\begin{equation}
    \mathbf{L}_0(\mathrm{curl},\Omega) = \{\mathbf{u} \in \mathbf{H}(\mathrm{curl},\Omega): \mathbf{n}\times\mathbf{u} = 0|_{\partial\Omega_\mathrm{D}} \}.
    \label{eq:L-space}
\end{equation}
In \eqref{eq:L-space}, $\mathbf{H}(\mathrm{curl},\Omega)$ is the space of square-integrable functions with square-integrable weak curl.
Note that the Neumann boundary integral arising in the deduction of \eqref{eq:weakFormulation} vanishes by applying the Neumann \gls{bc} $\mathbf{n}\cdot \left(\nu\left( \mathbf{b} \right)\mathbf{b}\right)=0$, where $\mathbf{n}$ is the outer normal unit vector. 
The Dirichlet \gls{bc} $\mathbf{n}\times\mathbf{a}=0$ is enforced in $\mathbf{L}_0(\mathrm{curl},\Omega)$. 

The magnetic vector potential is approximated within a finite element space as
\begin{equation}
    \mathbf{a}=\sum_{j=1}^{E} \hat{a}_{j} \mathbf{w}_j,
\end{equation}
where $\hat{a}_j$ are the \glspl{dof}, $E$ is the number of \glspl{dof}, and $\mathbf{w}_j$ denotes Nédélec basis functions of the first kind and the first order \cite{nedelec1980mixed}. We apply the Ritz-Galerkin procedure, such that the set of test functions is the same as the set of shape (basis) functions. 

\subsection{2D Magnet Model}
\begin{figure}[t]
    \centering
    \includegraphics[scale=0.4]{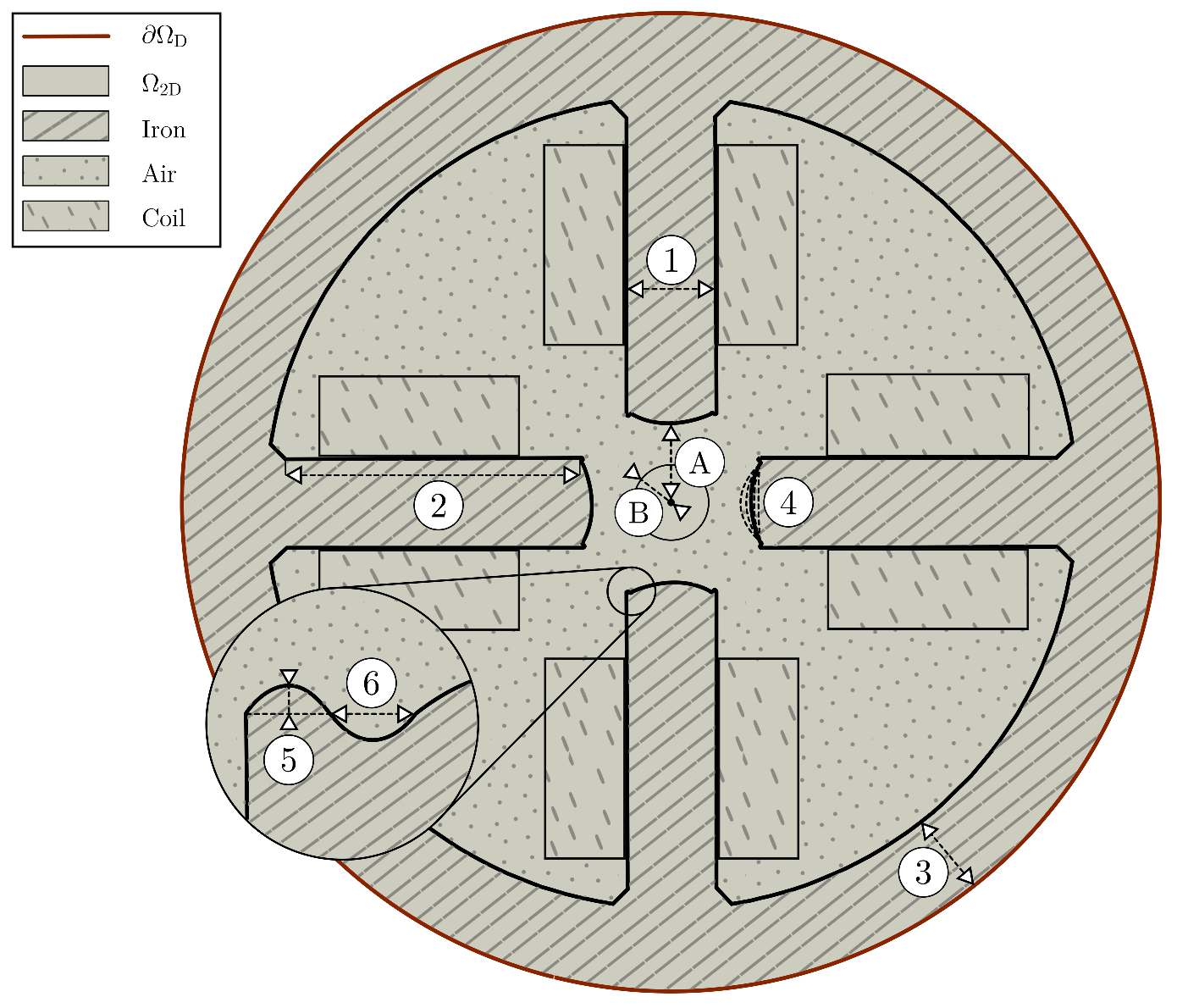}
    \caption{2D cross section of the quadrupole magnet. The computational domain, its boundaries, and the different material-based subdomains are presented as in the legend. The identifiers $1 - 6$ and the letters A, B, denote the geometrical parameters of the magnet, as described in Table \ref{tab:geometricalParams}.}
    \label{fig:fullModel1}
\end{figure}

The magnet model is further simplified taking into account the translation invariance of the magnet along the $z$-axis.
Therefore, we may consider a magnetic vector potential perpendicular to a $2$D cross section of the $3$D domain, in which case $\mathbf{a} = [0,0,a_z = a_z(x,y)]^\top$ and $\mathbf{j} = [0,0,j_z]^\top$.
We denote the computational domain of the 2D cross section of the quadrupole magnet with $\Omega_{\text{2D}}:=\{\mathbf{x}\in\mathbb{R}^2 \:\: | \:\: \left\lVert\mathbf{x}\right\rVert_2\leq R\}$, which corresponds to a circle with radius $R\in\mathbb{R}$ around the origin. Figure \ref{fig:fullModel1} depicts the full 2D quadrupole magnet model contained in $\Omega_{\text{2D}}$, where the corresponding geometry descriptors and numerical identifiers are given in Table \ref{tab:geometricalParams}. The Dirichlet \gls{bc} $\mathbf{n}\times\mathbf{a} = 0$ is imposed on the boundary $\partial\Omega_\mathrm{D}$, also shown in Figure \ref{fig:fullModel1}.  

\begin{table}[h]
\centering
\caption{Geometrical parameters of the quadrupole magnet.}
\begin{tabular}{lccccc} 
\toprule
 Description & Identifier & Notation & Value & Units \\ 
\midrule 
Pole width & \scalebox{0.75}{$\circled{1}$} & $x_1$ & $[14.0, 20.0]$ & \si{mm} \\
Pole height & \scalebox{0.75}{$\circled{2}$} & $x_2$ & $[45.0, 120.0]$ & \si{mm} \\
Yoke height & \scalebox{0.75}{$\circled{3}$} & $x_3$ & $[10.0, 25.0]$ & \si{mm} \\
Pole bending\tablefootnote{\textcolor{black}{The pole tip profile is modeled using a B-spline that is dependent on the design variable $x_4$. Only the radial coordinate of the resulting parametrization is proportional to $x_4$. Accordingly, in the case of the $\frac{1}{8}$-th quadrupole, $x\propto \mathrm{acosh}(cx_4)$.}} & \scalebox{0.75}{$\circled{4}$}  & $x_4$ & $[0.03, 0.1]$ & - \\
Shim height  & \scalebox{0.75}{$\circled{5}$} & $x_5$ & $[0.1, 0.55]$ & \si{mm} \\
Shim width  & \scalebox{0.75}{$\circled{6}$} & $x_6$ & $[0.4, 1.2]$ & \si{mm} \\
Bore radius & \scalebox{0.75}{$\circled{A}$} & - & $15.0$ & \si{mm} \\
Reference radius & \scalebox{0.75}{$\circled{B}$} & $r_\text{ref}$ & ${\color{black}11}$ & \si{mm} \\
\bottomrule
\end{tabular}
\label{tab:geometricalParams}
\end{table}

Table \ref{tab:geometricalParams} also presents the intervals $[a_i, b_i]$, within which the geometrical parameters are allowed to vary during the optimization procedure, i.e., $a_i \leq x_i \leq b_i$, $i=1,\dots,6$. 
Parameters with constant values throughout the optimization are denoted with the identifiers A and B. 
Shims are included in the quadrupole model as important pole adjustments, which can lead to significant improvements in the homogeneity of the magnetic field, thus, to field quality improvements as well \cite{shims}. 
\textcolor{black}{Further improvements can also be achieved through pole shape optimization as in \cite{LeBecIPAC2014}.}
The non-linear material of the yoke and the poles is modeled with a Brauer curve approximation \cite{BrauerCurve} upon $1010$-Steel and implemented in a FE solver with the Newton method \cite{getdp}. 
The model is  implemented using three open-source tools, namely, the mesh generator \texttt{Gmsh} \cite{gmsh}, the \texttt{GetDP} FE solver \cite{getdp}, and the \texttt{ONELAB} interface \cite{onelab}. 

\textcolor{black}{\subsection{Convergence of the Finite Element Discretization}}
\begin{figure}[t]
    \centering
    \includegraphics[scale=0.32]{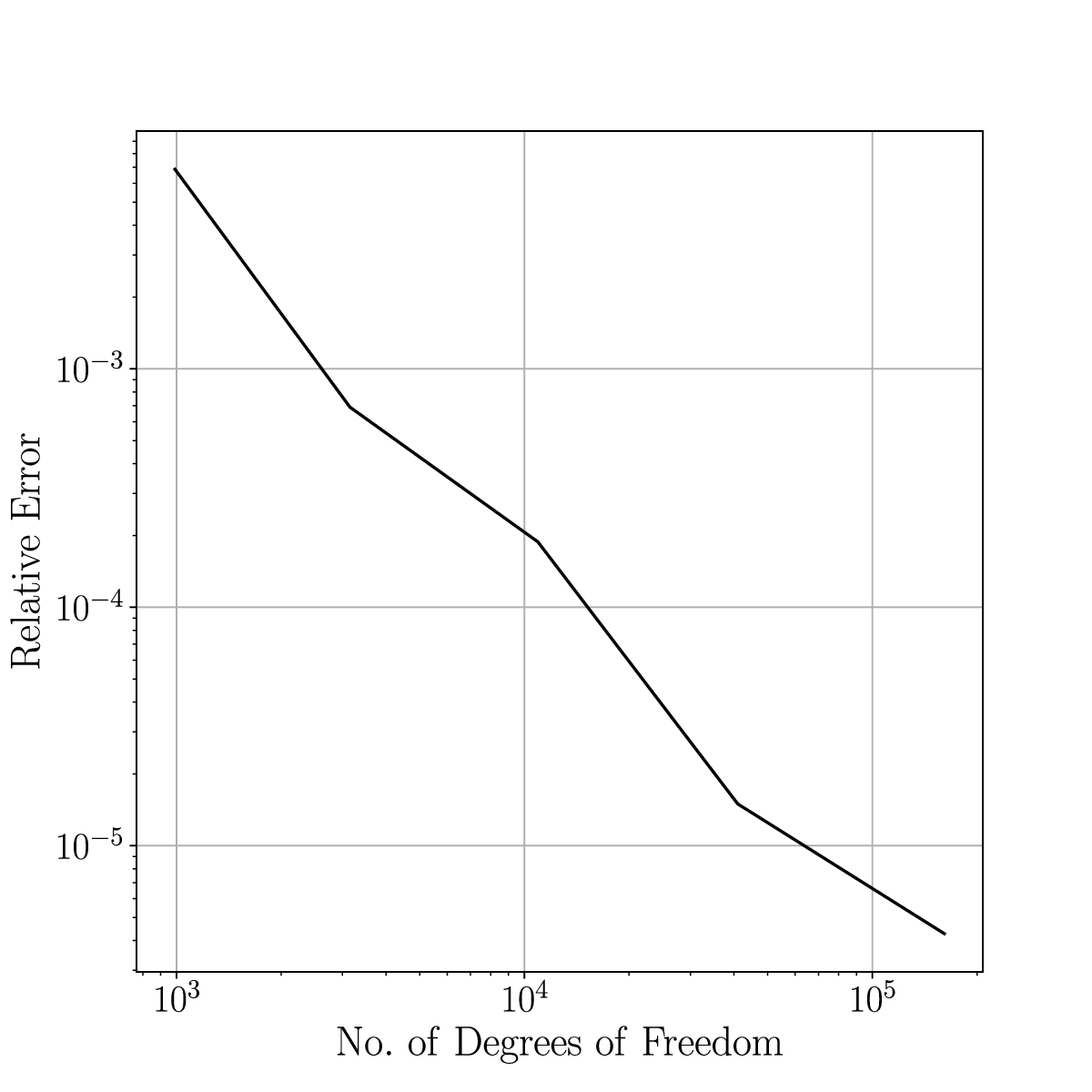}
    \caption{\textcolor{black}{Convergence of the FE discretization. Four refinements of the $2$D FE mesh have been executed, ranging from 990 dofs to 161028 dofs.}}
    \label{fig:convFE}
\end{figure}
\textcolor{black}{The convergence of the FE discretization of the magnet model is investigated by solving Equation \eqref{eq:MS1_3D} for a repeatedly finer $2$D mesh, while for each refinement step the magnetic energy in the computational domain $\Omega$ is post-processed.
Figure \ref{fig:convFE} depicts the resulting relative error of the magnetic energy over the number of dofs, indicating a first order polynomial convergence with respect to the number of $2$D mesh nodes. For the further results in this paper, the magnetostatic problem is solved with $18665$ $2$D mesh nodes and $9081$ triangular elements for each model evaluation of the quadrupole magnet.}

\subsection{Aperture Field Quality}
\label{sec:aperture_field_quality}
One of the most important quantities of interest to be taken into account during the design of a quadrupole magnet, is the field quality here represented by the harmonic distortion factor $Q$, which can be computed upon the multipole coefficients of the calculated field in the magnet's aperture. \textcolor{black}{The harmonic distortion factor captures the relationship between the desired multipoles such as the quadrupole components, and undesired multipoles such as duodecapole components and higher order terms. Therefore, $Q$ is a measure for the magnetic field quality of the quadrupole magnet and can be calculated as follows:} the FE solution $a_z$ is evaluated at a reference circle with radius $r_\text{ref}$. The result is then represented by a Fourier series with Fourier coefficients $a_p$ and $b_p$, such that
\begin{equation}
    a_z(r_\text{ref},\varphi) = \sum_{p=0}^{\infty}\left( a_p \cos{(p\varphi)}+ b_p \sin{(p\varphi)}\right),
\end{equation}
using the polar coordinate system $\left(r, \varphi\right)$.
Then, the Fourier characterization of the magnetic vector potential and the magnetic flux density in the beam aperture are given as \cite{russenschuck2011field}
\begin{align}
    a_z(r,\varphi) &= \sum_{p=0}^{\infty} \left( \frac{r}{r_\text{ref}}\right)^p \left( a_p \cos{(p\varphi)}+ b_p \sin{(p\varphi)}\right), \\
    b_r(r,\varphi) &= \sum_{p=1}^{\infty}\left( \frac{r}{r_\text{ref}}\right)^p  \frac{p}{r} \left( -a_p \sin{(p\varphi)}+ b_p \cos{(p\varphi)}\right), \\
    b_\varphi(r,\varphi) &= \sum_{p=1}^{\infty}\left( \frac{r}{r_\text{ref}}\right)^p  \frac{p}{r} \left( -a_p \cos{(p\varphi)}- b_p \sin{(p\varphi)}\right).
\end{align}
The evaluation at $r=r_\text{ref}$ for the radial magnetic flux density yields
\begin{equation}
    b_r(r_\text{ref},\varphi) = \sum_{p=1}^{\infty} \left( \underbrace{-a_p \frac{p}{r_\text{ref}}}_{B_p(r_\text{ref})} \sin{(p\varphi)} + \underbrace{b_p \frac{p}{r_\text{ref}}}_{A_p(r_\text{ref})} \cos{(p\varphi)}\right), 
\end{equation}
where $B_p$ and $A_p$ are called \emph{normal} and \emph{skew harmonic coefficients}, respectively. 
The harmonic distortion factor $Q(r_\mathrm{ref})$ in the aperture of a $2P$-pole magnet considered at a reference radius $r_\mathrm{ref}$ can be obtained from the harmonic Fourier coefficients using the formula \cite{russenschuck2011field}
\begin{equation}
    Q_P(r_\mathrm{ref}) = \frac{1}{{\color{black}A}^2_P(r_\mathrm{ref})} \sum_{\substack{ p=1 \\ p \ne P}}^{\infty} \left( B^2_p(r_\mathrm{ref}) + A^2_p(r_\mathrm{ref})\right).
\end{equation}
For a quadrupole magnet $\left(P=2\right)$, with a quadrupole field gradient
\begin{equation}
    g(r) = {\color{black}A}_2(r), \label{eq:quadrupoleGradient}
\end{equation}
the magnetic field should be close to a pure quadrupole field, i.e., the field gradient $g$ should be high in comparison to the other multipole coefficients ${\color{black}A}_p$, with $p=1,\dots ,C$, and $p\ne2$, where $C$ refers to the truncation coefficient. 
Accordingly, the corresponding harmonic distortion factor for a quadrupole magnet, given by
\begin{equation}
        Q_2(r_\mathrm{ref}) = \frac{1}{g^2(r_\mathrm{ref})} \sum_{\substack{ p=1 \\ p \ne 2}}^{C} \left(B^2_p(r_\mathrm{ref}) + A^2_p(r_\mathrm{ref})\right), 
        \label{eq:Q_quadrupole}
\end{equation}
should be in the order of $10^{-4}$ \cite{russenschuck2011field}. 

\section{Multi-Objective Optimization}\label{sec:Optimization}

In \gls{moo}, we seek to minimize (or maximize, depending on the problem at hand) a set of - possibly conflicting - objective functions $f_i(\mathbf{x}): \mathcal{X} \rightarrow \mathbb{R}$, $i = 1,...,k$, $k \geq 2$, where $\mathbf{x} = \left(x_1,\dots,x_n\right) \in \mathcal{X}$ denotes the vector of decision variables, equivalently, optimization parameters, and the domain $\mathcal{X} \subset \mathbb{R}^n$ is referred to as the \emph{feasible decision region} \cite{gunantara2018review, justesen2009multi}. 
For a feasible parameter vector $\mathbf{x} \in \mathcal{X}$, a corresponding feasible objective vector $\mathbf{z} =  \left(f_1(\mathbf{x}), \dots, f_k(\mathbf{x})\right)$ is obtained, where $\mathbf{z} \in \mathcal{Z}$ and $\mathcal{Z} \subset \mathbb{R}^k$ is called the \emph{feasible objective region}.

The structure of $\mathcal{X}$ is induced by a set of constraints applied to the decision variables. 
In the specific case of geometry optimization, these constraints are often given in the form of bounding box intervals, similar to the ones shown in Table~\ref{tab:geometricalParams}, such that $\mathcal{X} := [a_1, b_1]\otimes \cdots \otimes [a_n, b_n]$, where $a_i \leq x_i \leq b_i$, $i=1,\dots,n$. 
Given this set of constraints, the \gls{moo} problem reads
\begin{align}
\label{eq:min_prob}
\min_{\mathbf{x} \in \mathcal{X}} \left\{f_1(\mathbf{x}),...,f_k(\mathbf{x}) \right\}.
\end{align}

In most cases, a parameter vector that minimizes all objective functions simultaneously does not exist.
It is therefore necessary to have a method of comparing a set of solutions while taking into account the satisfaction of all objectives.
This issue is resolved using the concept of \emph{dominating solutions}. 
Assuming two parameter vectors $\mathbf{x}, \tilde{\mathbf{x}} \in \mathcal{X}$ arising in a minimization procedure, then $\tilde{\mathbf{x}}$ dominates $\mathbf{x}$ if
\begin{align}
\label{eq:dominance}
      f_i(\tilde{\mathbf{x}}) < f_i(\mathbf{x})  \text{ and } f_j(\tilde{\mathbf{x}}) \leq f_j(\mathbf{x}), \forall i,j \in \{1,...,n\},\, i\neq j.
\end{align}
A parameter vector that cannot be dominated is called \emph{Pareto-optimal} \cite{luc2008pareto}. 
In essence, Pareto-optimality means that the current solution cannot be further improved with respect to one of the objectives, without simultaneously deteriorating another objective. 
The set of Pareto-optimal solutions is referred to as the \emph{Pareto front}. 
A Pareto front is bounded by the \emph{ideal} and \emph{nadir} objective vectors, respectively denoted with $\mathbf{z^*}$ and $\mathbf{z}^{\text{nad}}$. 
The former is obtained by individually optimizing the objective functions and the latter by approximating the worst objective values of the Pareto front.

\subsection{Genetic Algorithms}
\glspl{ga} belong to a class of population based, stochastic optimization algorithms  which solve optimization problems by only allowing candidate solutions with a promising ``gene pool'' to reproduce and propagate through generations $t=1,2,\dots , T$, where $T$ is the final generation \cite{schmitt2001theory, weile1997genetic}. This generation-based evolution of candidates is realized by creating a sequence of subsets $\mathcal{P}$ in the feasible decision space called \emph{populations}, which eventually converge to a set of minimizers. 
In that way, a sequence  $(\mathcal{P}_t)_{t\in \mathbb{N}} \subset \mathcal{X}$ with 
\begin{equation}
    \mathcal{P}_t \xrightarrow{t\rightarrow \infty}\mathcal{P^*} \label{eq:theoreticalLimitGenerations}
\end{equation} 
is generated, such that each $\mathbf{x}\in \mathcal{P^*}$ is a minimizer of \eqref{eq:min_prob}. To deal with the limitations of realistic and thus finite calculations, Equation \eqref{eq:theoreticalLimitGenerations} needs to be truncated by a final generation $T$, so that with
\begin{equation}
    \mathcal{P}_t \xrightarrow{t\rightarrow T}\mathcal{P}_T, \label{eq:practicalLimitGenerations}
\end{equation}
a set of generation-related minimizers $\mathbf{x}\in\mathcal{P}_T$ is considered. 

The sequence of populations $\left(\mathcal{P}_t\right)_{t=1,2,\dots,T}$ is dictated by the genetic operators \emph{crossover} $C_{\sigma}$, \emph{mutation} $M_r$, and \emph{fitness selection} $F$, such that
\begin{align}
    \mathcal{P}_{t+1}=F\circ M_m \circ C_{\sigma}\left(\mathcal{P}_t\right),
\end{align}
where $\circ$ expresses the concatenation of the sequentially applied genetic operators on the current population $\mathcal{P}_t$. 
The indices $\sigma$ and $m$ respectively denote the crossover and mutation rates of the corresponding operators, where $\sigma, m \in [0,1]$. 
The crossover and mutation rates determine the probability that the given operator is applied to a given population. 
The crossover operator $C_{\sigma}$ describes how sample solutions are recombined to generate new solutions for the next population. 
The mutation operator $M_m$ describes random distortions to the elements of a population and is particularly significant for the convergence of the \gls{ga}, as it ensures that the objective space $\mathcal{X}$ is searched comprehensively and that the limit $\mathcal{P^*}$ is initialization-independent. 
Last, the fitness selection operator $F$ allocates a  fitness value to the population members and selects those with the highest values to progress to the next generation.

In this work, simulated binary crossover \cite{SimulatedBinaryCrossover1AndMutation1, SimulatedBinaryCrossover2} and polynomial mutation \cite{golchha2015non, yusoff2011overview} are employed as crossover and mutation operators, respectively. 
For fitness evaluation, we use the \gls{fe} model of the quadrupole to compute the magnetic field distribution in the magnet and evaluate the aperture field quality, which in turn  determines the fitness of a given population.  
As a selection operator, we use binary tournament selection \cite{Miller1995GeneticAT}, due to its ease of implementation and robustness against stochastic noise.

\subsection{Non-Dominated Sorting Genetic Algorithms \& NSGA-III}
As noted before, \glspl{ga} were originally developed to solve single-objective optimization problems. 
Therefore, they cannot address a number of issues related to \gls{moo}, such as dealing with multiple objective functions and ensuring diversity in the populations. 
These issues have been addressed with the introduction of \glspl{nsga} \cite{NSGAI, NSGAII, NSGAIII, jain2013evolutionary}. 
In this work, we resort to the so-called \gls{nsga}-III algorithm \cite{NSGAIII, jain2013evolutionary}, which is briefly discussed in the following.

First, \gls{nsga}-III deals with the issue of optimizing numerous objective functions by preferentially handling solutions that dominate other members of a population according to definition \eqref{eq:dominance}, by using \emph{non-dominated sorting}. 
Given the current population $\mathcal{P}$, non-domination sorting partitions the population $\mathcal{P}=\mathcal{F}_1\cup\cdots\cup\mathcal{F}_{N-1}$ into disjoint sets $\mathcal{F}_i$, which form the hierarchy $\mathcal{F}_1<\mathcal{F}_2<\cdots<\mathcal{F}_{N-1}$. 
This hierarchy is induced according to a domination factor $n_p=0,1,...,N-1$, which indicates how many solutions from an equal or lower hierarchy level dominate the given solution. 
The hierarchy construction is as follows: The set $\mathcal{F}_1$ includes $\mathbf{x}\in \mathcal{P}$, which have $n_p = 0$, i.e, they are not dominated by any solution. 
Then, we consider the set $\mathcal{Q} = \mathcal{P} \backslash\mathcal{F}_2$ and decrement the domination factor for $n_p\rightarrow n_p-1$ for all $q\in \mathcal{Q}$ that are dominated by an element of $\mathcal{F}_1$. 
This process is repeated iteratively, thus yielding a hierarchical set sequence. 
The sets with lower domination factors qualify to the next population, whereas the sets with higher domination factors are discarded. 
It is thus ensured that that the solutions propagating to future generations are Pareto-optimal with respect to the current population they belong to.

To ensure population diversity, \gls{nsga}-III adds another operation to the fitness selection procedure. 
Therein, the objective vectors $\mathbf{z}$ are normalized to the unit cube by using the ideal objective vector $\mathbf{z}^*$ and the nadir objective vector $\mathbf{z}^{\text{nad.}}$. 
In that way, it is possible to consider objective functions that are scaled differently. 
Then, reference points on the unit hypercube are chosen, which lie on a simplex \cite{DasAndDennis}. The reference points typically have a space-filling property and the objective vectors are projected to the reference points. 
The population members are then determined by an explicit diversity-preserving mechanism.

\section{Quadrupole Magnet Optimization}
\label{sec:magnet-optimization}
The \gls{moo} concerns maximizing the absolute value of the field gradient $g$, as introduced in Section \ref{sec:aperture_field_quality}, while minimizing the outer radius $R$ of the magnet. The optimization parameters are the six geometrical parameters listed in Table~\ref{tab:geometricalParams} and the current density $\mathbf{j}$. 
The latter takes values within the interval $[1.0, 20.0]$~\si{A/mm^2} and in the following is denoted with $x_7$. 
Then, the \gls{moo} problem reads
\begin{align}
\label{MOO:obj}
\min_{\mathbf{x} \in \mathcal{X}} \left\{-|g(\mathbf{x})|, \, R(\mathbf{x}) \right\}, 
´\end{align}
where the feasible decision space is $\mathcal{X}=[a_1, b_1]\otimes \cdots \otimes [a_7, b_7]$.

\begin{figure}[H]
    \centering
    \includegraphics[scale=0.42]{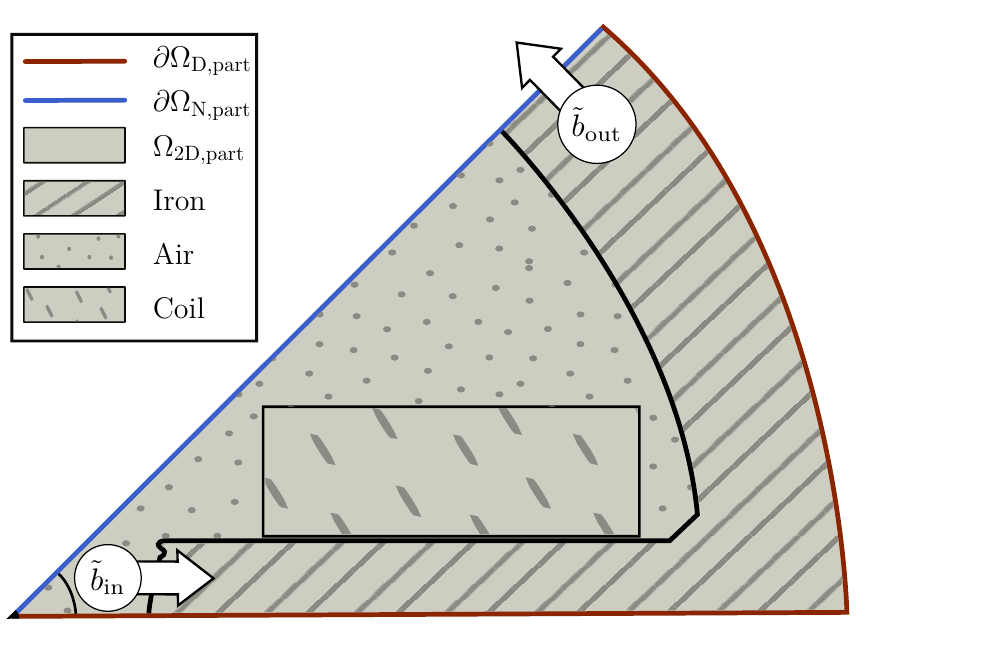}
    \caption{One-eighth of the 2D quadrupole magnet model.}
    \label{fig:partModel1}
\end{figure}

Instead of optimizing the full 2D magnet model shown in Figure \ref{fig:fullModel1}, we exploit the three mirror symmetries of the magnet model in order to reduce it to the one-eighth segment depicted in Figure \ref{fig:partModel1}. 
This model reduction leads to a significant improvement in terms of the computational cost of the \gls{fem}. 

Besides constraints on the geometrical parameters and the current density, we also introduce constraints on the absolute duodecapole gradient $g_\text{d}$, as well as on the saturation behavior of the iron yoke of the magnet, which read
\begin{subequations}
\label{MOO:Constraints}
\begin{align}
 |g_\text{d}(\mathbf{x})|    &\leq |g(\mathbf{x})|  10^{-2},  \\
\tilde{b}_\text{in} &\leq b_\text{th}, \\
\tilde{b}_\text{out}  &\leq b_\text{th}.
\end{align}
\end{subequations}
Analogously to the quadrupole field gradient $g$ in Equation \eqref{eq:quadrupoleGradient}, the duodecapole gradient is given by $g_\text{d} = B_6(r_\text{ref})$. In any case,the additional constraints demand that the duodecapole gradient $g_\text{d}$ should not exceed a certain percentage of the absolute field gradient $g$, as well as that the magnetic flux density entering the pole nose $\tilde{b}_\text{in}$ and the magnetic flux density exiting the yoke bend $\tilde{b}_\text{out}$ are bounded by a fixed saturation threshold value $b_\text{th}$. 
The magnetic flux densities $\tilde{b}_\text{in}$ and $\tilde{b}_\text{out}$  are illustrated in Figure \ref{fig:partModel1}. 
Both are obtained by post-processing the FE solution $a_z$.

The solutions to the MOO problem defined by equations \eqref{MOO:obj}-\eqref{MOO:Constraints} are obtained, as previously noted, using the \gls{nsga}-III algorithm. 
In particular, the implementation of the algorithm which is available in the open-source, Python-based optimization software \texttt{pymoo} is used \cite{pymoo}.

\section{Numerical Results}\label{sec:Results}
In this section, the numerical results of the MOO are presented, which are additionally employed to identify the best solutions, i.e., the ones corresponding to the most suitable magnet designs. The notion of the best solution is split into:
\begin{itemize}
    \item \emph{Best balanced solution}, that is, the Pareto-optimal solution with the lowest Euclidean distance to the ideal objective vector, where the latter is approximated with respect to the final Pareto front.
    \item \emph{Best field gradient solution}, that is, the Pareto-optimal solution found in the final Pareto front, which results in the highest field gradient.
\end{itemize}

\begin{figure}[t!]
\centering
\includegraphics[scale=0.38]{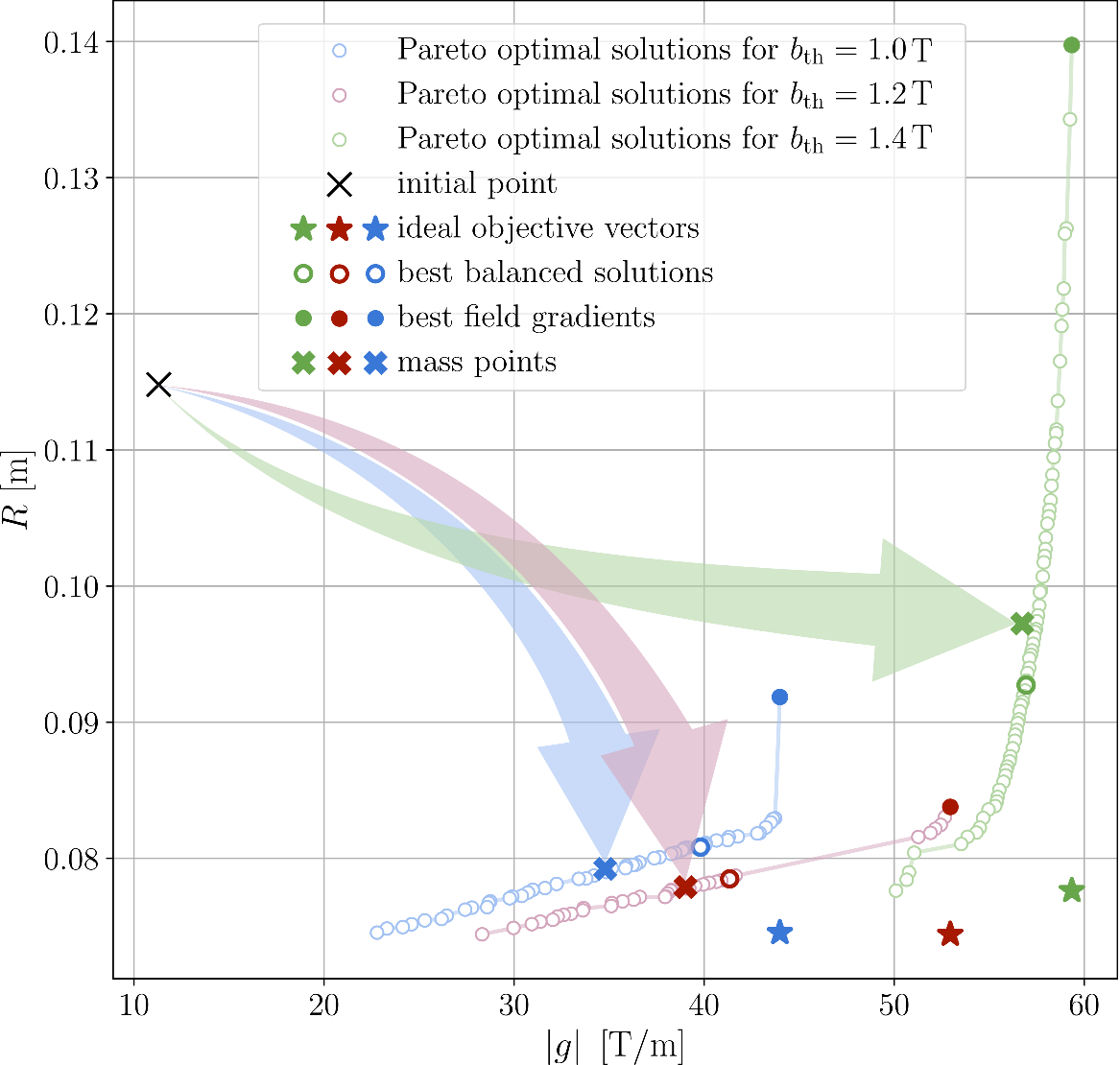}
\caption{{\color{black}MOO results for three optimization runs, each corresponding to a different saturation threshold $b_\text{th}$. 
The arrows show the Pareto front movement from the initial point ({\small$\times$}) towards the mass point of the final Pareto front ($\,${\scriptsize \ding{54}}$\,$).}} 
\label{fig:pf_comparison}
\end{figure}

Using \gls{nsga}-III with a crossover rate $\sigma = 1.0$ and a polynomial mutation rate $m = 0.1$, each optimization run is performed for $T=300$ generations, with an initial population of $80$ individuals, and an offspring population of $56$ individuals. 
These empirically enforced values remain fixed during several optimization runs, where the influence of different saturation behaviors of the iron are investigated. 
To analyze the saturation behavior, three optimization runs are performed, within which the saturation threshold $b_\text{th}$, which is embedded in the optimization's constraints as in Equations \eqref{MOO:Constraints}, is selected as \SI{1.0}{\tesla}, \SI{1.2}{\tesla}, or \SI{1.4}{\tesla}, respectively. 
Each optimization run is performed upon the same feasible decision space $\mathcal{X}$. 
The initial point is chosen to be the center point of $\mathcal{X}$ and represents a naive but admissible choice of geometry with respect to the optimization constraints, see Figure \ref{fig:pf_comparison}. 
The associated initial objective values are $R=\SI{0.115}{\meter}$ and $|g|=\SI{11.305}{\tesla / \meter}$. 
Here, the initial point serves as a reference point to compare the locations of the different final Pareto fronts. 

Figure \ref{fig:pf_comparison} shows the Pareto front movement and the location of the final Pareto front after $T=300$ generations for the three chosen saturation thresholds. 
Therein, the Pareto front movement is obtained by connecting the mass points of the fronts obtained from each generation of the \gls{nsga}-III, where the mass point is defined as the mean of all Pareto optimal solutions.
As can be seen, both the final Pareto front and the front movement depend strongly on the chosen saturation threshold. 
\textcolor{black}{For $b_\text{th} = \SI{1.0}{\tesla}$ and $b_\text{th} = \SI{1.2}{\tesla}$, the final Pareto fronts are biased towards low values of the outer radius $R$.}
In contrast, the Pareto front for $b_\text{th} = \SI{1.4}{\tesla}$ is biased towards high field gradient values. 
As for the individual evolution of the different objective spaces, Figure \ref{fig:pf_movement_appendix} in the Appendix shows each of the optimization solutions and their Pareto front movements individually.

\begin{figure}[t!]
    \centering
    \includegraphics[scale=0.4]{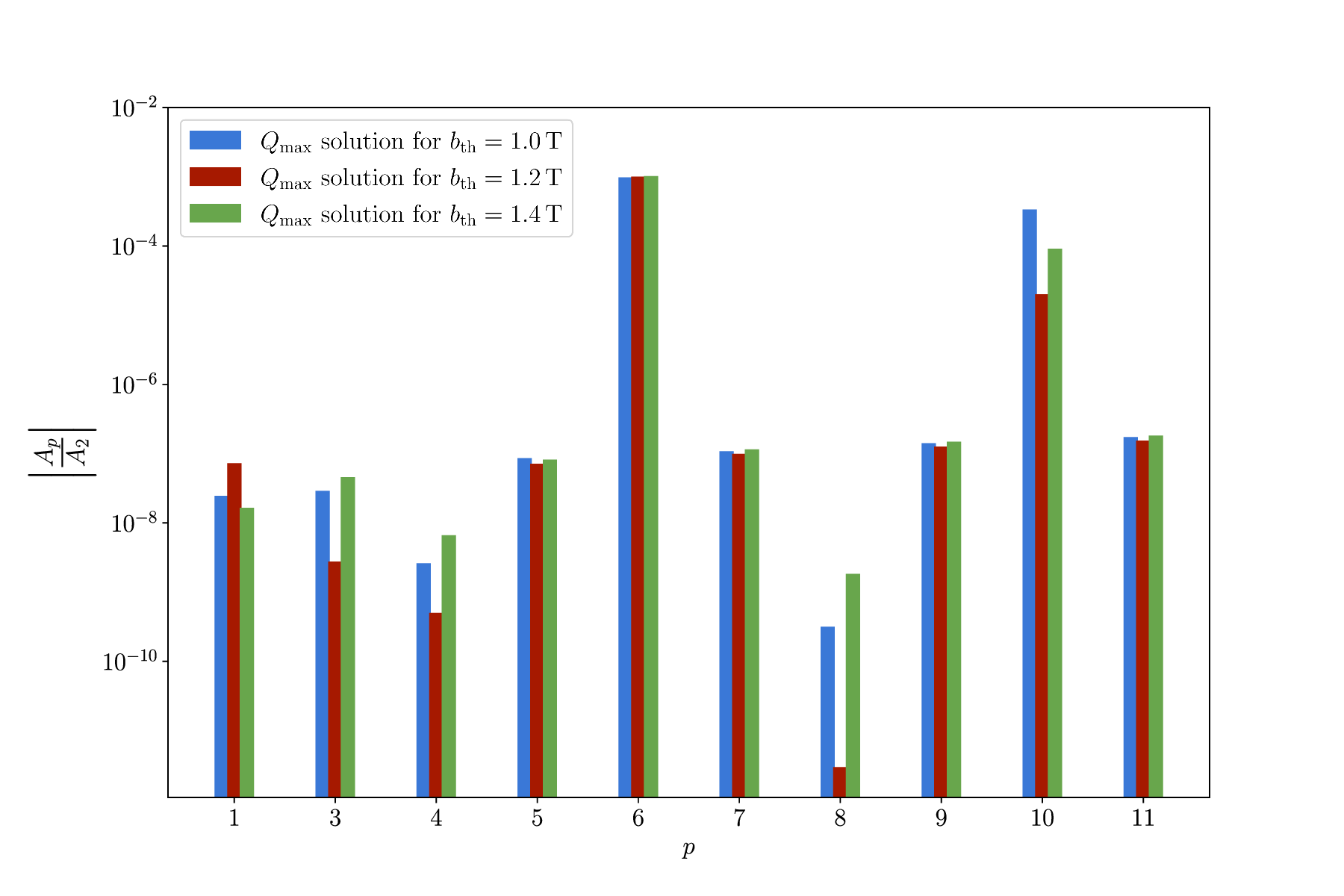}
    \caption{\textcolor{black}{Relative skew harmonic coefficients of the Pareto front solutions associated with the highest, i.e., worst harmonic distortion factor for each saturation threshold $b_\text{th}$, respectively.}}
    \label{fig:multipoleCoeffsWorstCase}
\end{figure}

\textcolor{black}{Figure \ref{fig:multipoleCoeffsWorstCase} depicts the relative skew harmonic coefficients with respect to the quadrupole main component, of the Pareto front solutions associated with the highest, i.e., worst harmonic distortion factor $Q_\text{max}$ for each saturation threshold $b_\text{th}$, respectively. Further, it is verified that the Pareto optimal solutions of all optimization runs have a harmonic distortion factor $Q$ below or of order $10^{-4}$.} This observation is consistent with the theoretical requirements for the harmonic distortion factor of a  sufficiently undistorted quadrupole field, as mentioned in Section \ref{sec:aperture_field_quality}.


%

%

\begin{figure*}[t!]
    \captionsetup[subfigure]{justification=raggedright}
    \begin{subfigure}{0.1\linewidth}
        \caption{Best balanced solutions.}
        \includegraphics[scale=0.01]{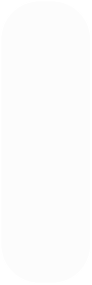}
        \label{fig:balancedSolutions}
    \end{subfigure}
    \begin{subfigure}{0.8\linewidth}
        \centering
        \includegraphics[scale=0.26]{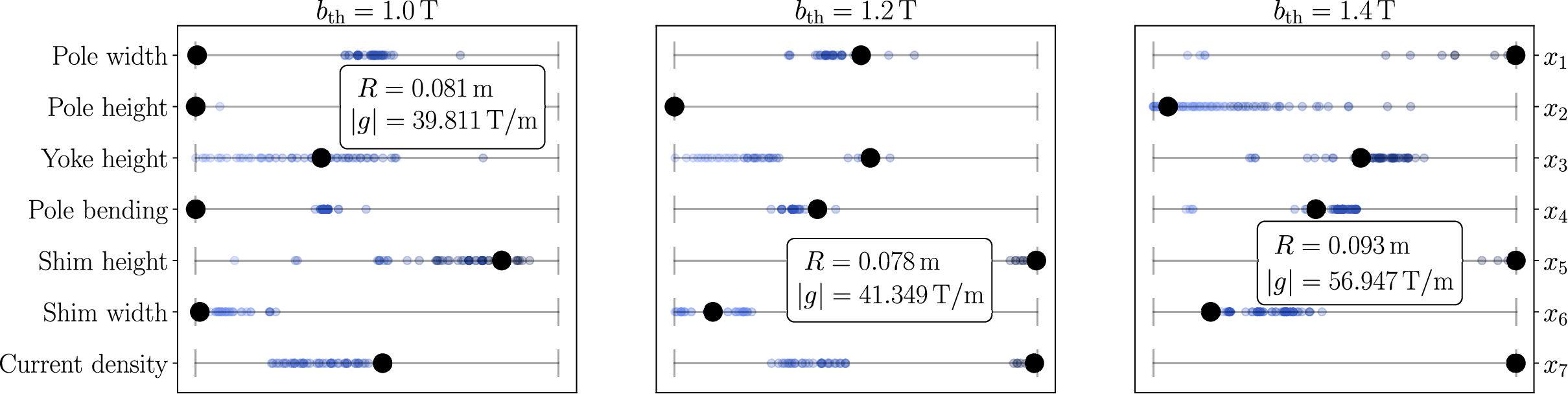}
    \end{subfigure}

    \begin{subfigure}{0.1\linewidth}
        \caption{Best field gradient solutions.}
        \includegraphics[scale=0.01]{Figures/whiteBox.pdf}
        \label{fig:bestFieldGradient}
    \end{subfigure}
    \begin{subfigure}{0.8\linewidth}
        \centering
        \includegraphics[scale=0.26]{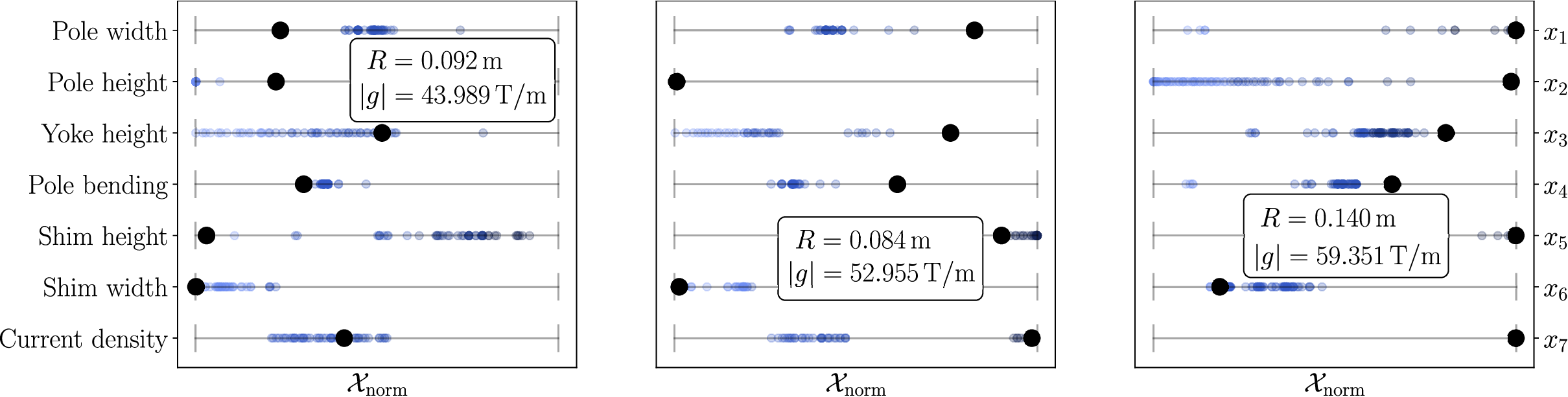}
    \end{subfigure}
\caption{{\color{black}Best solutions (normalized) for different saturation threshold values $b_\text{th}$. Additionally to the best solutions, the corresponding objective values $R$ and $|g|$ are given. Filled black circles: normalized parameter combinations of the best solutions. Transparent blue-colored circles: normalized parameter combinations of all final Pareto-optimal solutions with respect to the different saturation thresholds.}}
\label{fig:compareBalanceBestGrad}
\end{figure*}

Next, focus is shifted to the best solutions of the final Pareto front for each saturation threshold. 
For a better visual comparison of the best solutions, the parameter  combination corresponding to each solution is normalized to the unit cube $\mathcal{X}_\text{norm} = [0, 1]^7$. 
Using this normalization, the parameter distributions and the associated objective values of the best solutions are depicted in Figure \ref{fig:compareBalanceBestGrad}. 
As can be observed, the saturation threshold has a strong influence on the magnet's pole width, both for best balanced and for best field gradient solutions. 
This is attributed to the fact that higher saturation thresholds lead to a larger pole width design. 
It is further observed that larger pole widths lead to higher field gradients for both cases of best solutions. 
The saturation threshold has also a slight to moderate influence on the pole bending design. 
With regard to the current flow, it can be seen that, for both best solution cases, higher saturation thresholds allow higher current densities, as expected. 
In the case of the best balanced solutions (Figure \ref{fig:balancedSolutions}), it is observed that changes in the saturation threshold have almost no impact on the dimensions of the pole height, yoke height, and the shim geometry. 
Contrarily, the dimension of the pole height is more important in the case of the best field gradient solution, as can be seen in Figure \ref{fig:bestFieldGradient}. 
Additionally, the choice of the shim geometry seems to be more important for best field gradient solutions than for best balanced solutions.

\clearpage
\section{Conclusion} \label{sec:ConclusionAndOutlook}
This work presented a framework for optimizing predominantly the geometrical and secondarily the operational characteristics of a quadrupole magnet. 
The framework employs a \gls{moo} formulation, where two conflicting objectives must be satisfied, namely, high magnetic field quality and acceptable production cost. 
The \gls{moo} problem is solved by means of the so-called \gls{nsga}-III algorithm, which is a \gls{ga} suitably modified to address the issues arising in \gls{moo}. 
Therein, a magnetostatic \gls{fe} model of the magnet is employed in order to assess the quality of the magnetic field in the aperture of the magnet. 
Finally, the \gls{moo} problem is complemented with additional constraints on the duodecapole gradient and the saturation threshold of iron.

The numerical results indicate that saturation has a major impact on the obtained Pareto-optimal solutions.
Further analyzing the connection between optimization parameters and Pareto fronts, it is possible to deduce useful information regarding the impact of the different geometrical characteristics of the magnet onto the optimization objectives.
Importantly, all identified optimal magnet designs lead to a sufficiently low harmonic distortion factor, while it is possible to identify designs with an acceptable balance between field quality and production cost.

Future work should consider the utilization of a 3D magnet model, in order to consider fringe field effects during \gls{moo} studies.
The impact of eddy currents and power losses should also be taken into account.
As a result, \gls{moo} problems with $k>2$ objectives can be formulated and investigated, leading to even more improved magnet designs. 
To limit the computational burden of \gls{moo}, especially for an increased number of objectives, surrogate modeling approaches  could be considered \cite{loukrezis2022power}.
\section*{Declarations}
\textbf{Conflict of interest:} The authors declared no potential conflict of interests with respect to the research, authorship and/or publication of this
article.
\section*{Acknowledgments}
This work is supported by the Graduate School Computational Engineering within the Centre for Computational Engineering at the Technische Universit\"at Darmstadt. M.~von Tresckow acknowledges the support of the German Federal Ministry for Education and Research (BMBF) via the research contract 05K19RDB.
D.~Loukrezis and H.~De Gersem acknowledge the support of the German Research Foundation (DFG) via the research grant TRR 361 (grant number: 492661287).

\bibliographystyle{unsrt}

\bibliography{references}

%
\begin{appendices}
\section{Pareto Front Movements} \label{sec:Appendix}
\begin{figure}[H]
    \captionsetup[subfigure]{justification=raggedright}
    \hspace*{1cm} 
    \begin{subfigure}{.25\linewidth}
        \centering
        \includegraphics[scale=0.22]{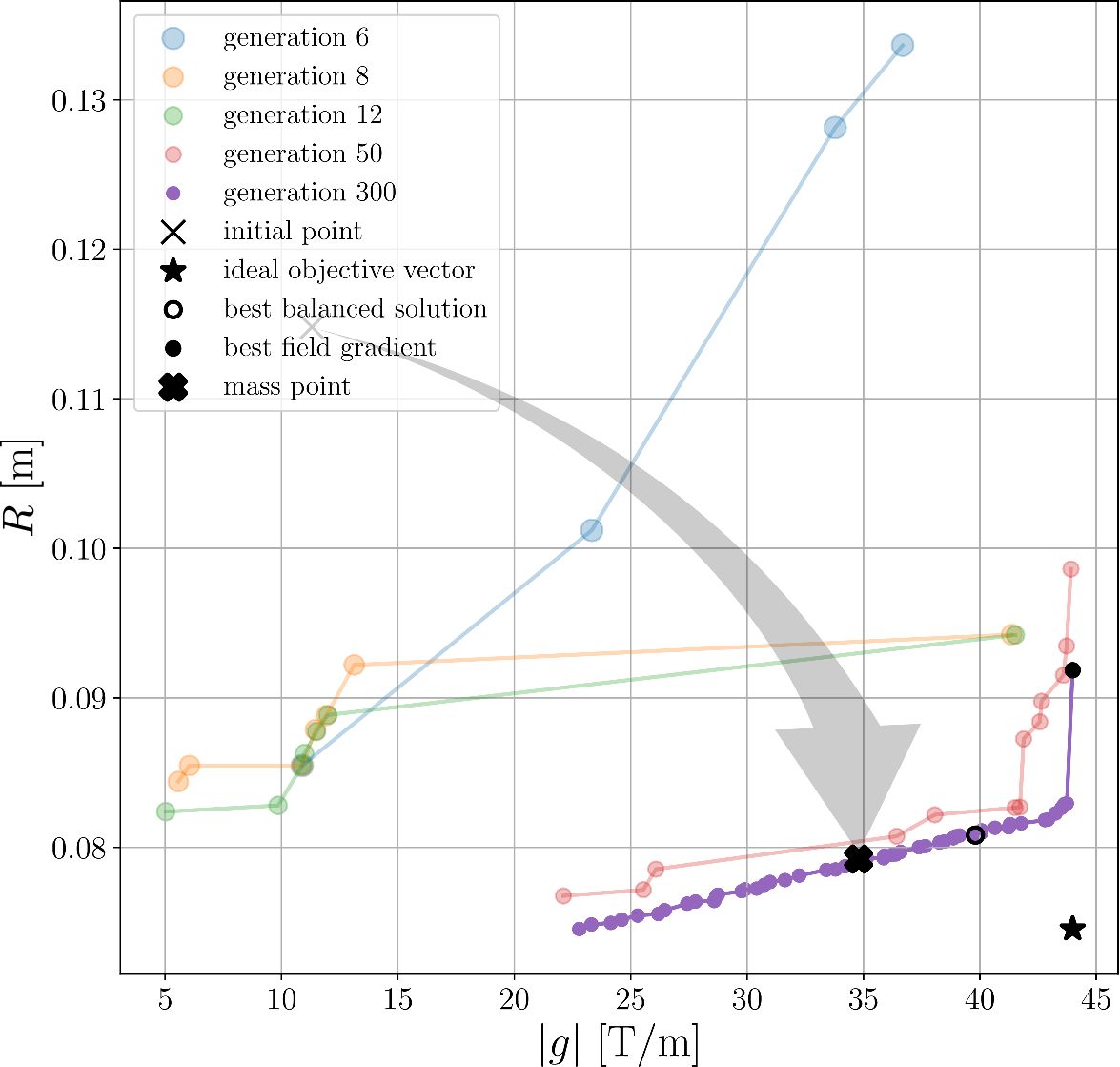}
        \caption{{\color{black}\small $b_\text{th} = \SI{1.0}{\tesla}$.}}
        \label{fig:pf_bsat1.2_appendix}
        \vspace*{0.25cm}
    \end{subfigure}\\%
    \vspace*{0.25cm}
    \hspace*{1cm}  
    \begin{subfigure}{.25\linewidth}
        \centering
        \includegraphics[scale=0.22]{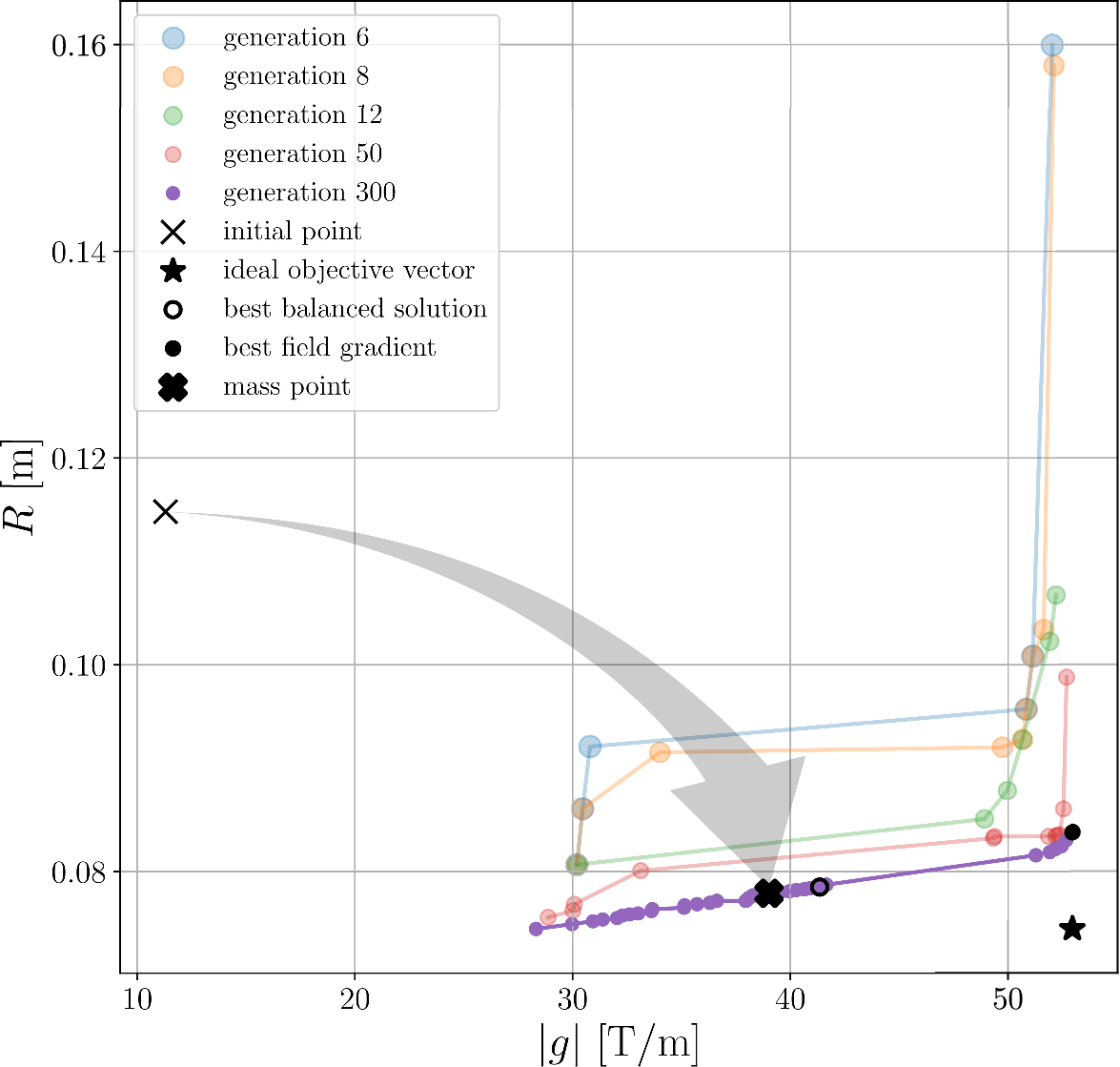}
        \caption{{\color{black}\small $b_\text{th} = \SI{1.2}{\tesla}$.}}
        \label{fig:pf_bsat1.0_appendix}
    \end{subfigure} \\
    \hspace*{1.1cm}  
    \begin{subfigure}{.25\linewidth}
        \centering
        \includegraphics[scale=0.22]{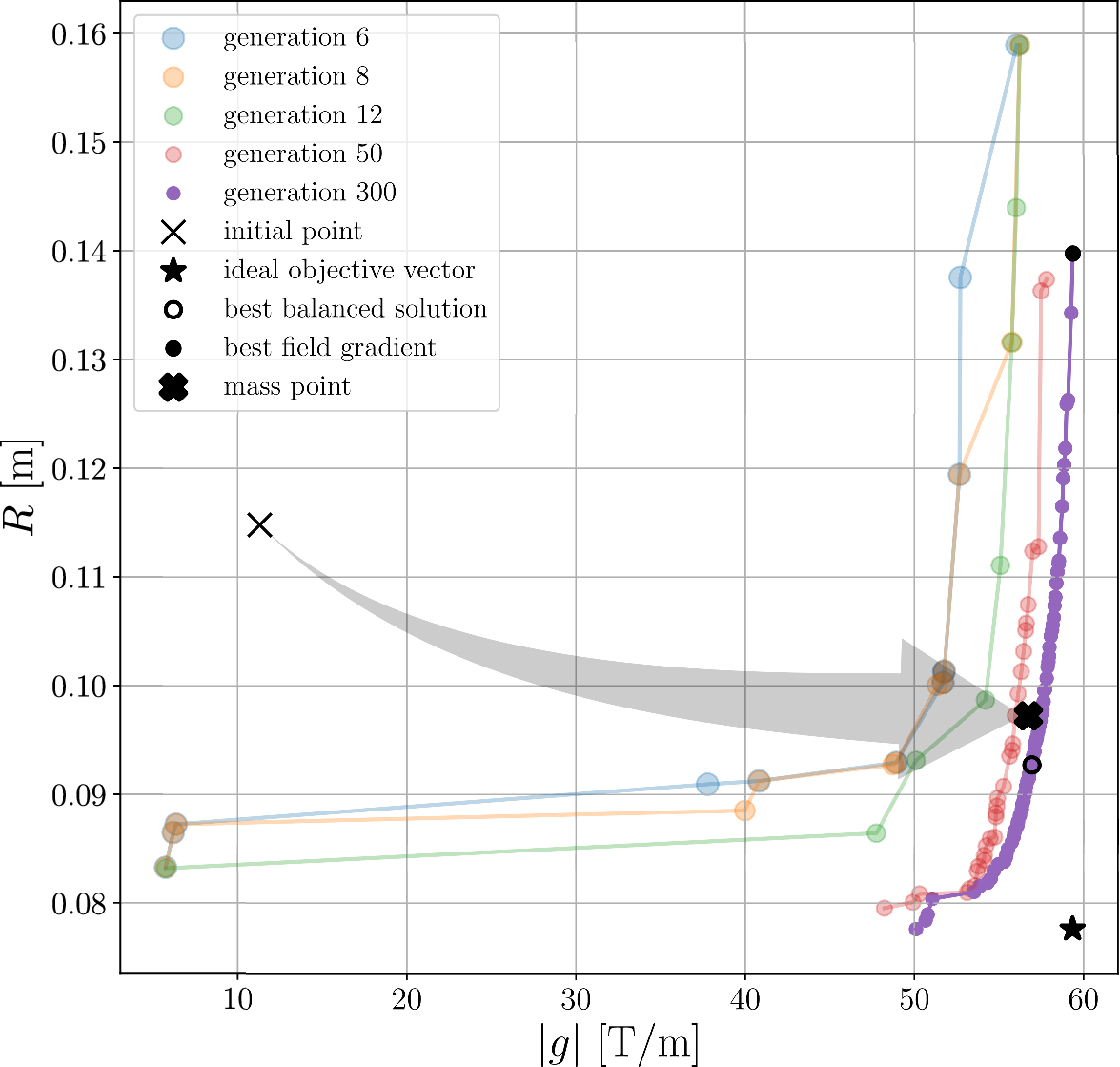}
        \caption{{\color{black}\small $b_\text{th} = \SI{1.4}{\tesla}$.}}
        \label{fig:pf_bsat1.4_appendix}
    \end{subfigure}%
\caption{Pareto front movements for each saturation threshold. Colored points: Pareto fronts for different optimization generations. Grey arrow: Pareto front movement from the initial point ({\small$\times$}) to the mass point of the final Pareto front ($\,${\scriptsize \ding{54}}$\,$). 
}
\label{fig:pf_movement_appendix}
\end{figure}
\end{appendices}

\end{document}